\documentstyle[12pt,amssymbols]{article}
\def\journal{\topmargin .3in    \oddsidemargin .5in
        \headheight 0pt \headsep 0pt
        \textwidth 5.625in 
        \textheight 8.25in 
        \marginparwidth 1.5in
        \parindent 2em
        \parskip .5ex plus .1ex         \jot = 1.5ex}
\journal

\catcode`\@=11
\def\marginnote#1{}
\newcount\hour
\newcount\minute
\newtoks\amorpm
\hour=\time\divide\hour by60
\minute=\time{\multiply\hour by60 \global\advance\minute by-\hour}
\edef\standardtime{{\ifnum\hour<12 \global\amorpm={am}%
        \else\global\amorpm={pm}\advance\hour by-12 \fi
        \ifnum\hour=0 \hour=12 \fi
        \number\hour:\ifnum\minute<10 0\fi\number\minute\the\amorpm}}
\edef\militarytime{\number\hour:\ifnum\minute<10 0\fi\number\minute}
\def\draftlabel#1{{\@bsphack\if@filesw {\let\thepage\relax
   \xdef\@gtempa{\write\@auxout{\string
      \newlabel{#1}{{\@currentlabel}{\thepage}}}}}\@gtempa
   \if@nobreak \ifvmode\nobreak\fi\fi\fi\@esphack}
        \gdef\@eqnlabel{#1}}
\def\@eqnlabel{}
\def\@vacuum{}
\def\draftmarginnote#1{\marginpar{\raggedright\scriptsize\tt#1}}
\def\draft{\oddsidemargin -.5truein
        \def\@oddfoot{\sl preliminary draft \hfil
        \rm\thepage\hfil\sl\today\quad\militarytime}
        \let\@evenfoot\@oddfoot \overfullrule 3pt
        \let\label=\draftlabel
        \let\marginnote=\draftmarginnote
   \def\@eqnnum{(\theequation)\rlap{\kern\marginparsep\tt\@eqnlabel}%
\global\let\@eqnlabel\@vacuum}  }
\def\preprint{\twocolumn\sloppy\flushbottom\parindent 2em
        \leftmargini 2em\leftmarginv .5em\leftmarginvi .5em
        \oddsidemargin -.5in    \evensidemargin -.5in
        \columnsep .4in \footheight 0pt
        \textwidth 10in \topmargin  -.4in
        \headheight 12pt \topskip .4in
        \textheight 7.1in \footskip 0pt
        \def\@oddhead{\thepage\hfil\addtocounter{page}{1}\thepage}
        \let\@evenhead\@oddhead \def\@oddfoot{} \def\@evenfoot{} }
\def\numberbysection{\@addtoreset{equation}{section}
        \def\theequation{\thesection.\arabic{equation}}}
\def\underline#1{\relax\ifmmode\@@underline#1\else
        $\@@underline{\hbox{#1}}$\relax\fi}
\def\titlepage{\@restonecolfalse\if@twocolumn\@restonecoltrue\onecolumn
     \else \newpage \fi \thispagestyle{empty}\c@page\z@
        \def\thefootnote{\fnsymbol{footnote}} }
\def\endtitlepage{\if@restonecol\twocolumn \else \newpage \fi
        \def\thefootnote{\arabic{footnote}}
        \setcounter{footnote}{0}}  
\catcode`@=12
\relax
\def\figcap{\section*{Figure Captions\markboth
        {FIGURECAPTIONS}{FIGURECAPTIONS}}\list
        {Figure \arabic{enumi}:\hfill}{\settowidth\labelwidth{Figure 999:}
        \leftmargin\labelwidth
        \advance\leftmargin\labelsep\usecounter{enumi}}}
 \relax
\def\tablecap{\section*{Table Captions\markboth
        {TABLECAPTIONS}{TABLECAPTIONS}}\list
        {Table \arabic{enumi}:\hfill}{\settowidth\labelwidth{Table 999:}
        \leftmargin\labelwidth
        \advance\leftmargin\labelsep\usecounter{enumi}}}
 \relax
\def\reflist{\section*{References\markboth
        {REFLIST}{REFLIST}}\list
        {[\arabic{enumi}]\hfill}{\settowidth\labelwidth{[999]}
        \leftmargin\labelwidth
        \advance\leftmargin\labelsep\usecounter{enumi}}}
 \relax
\makeatletter
\newcounter{pubctr}
\def\publist{\@ifnextchar[{\@publist}{\@@publist}}
\def\@publist[#1]{\list
        {[\arabic{pubctr}]\hfill}{\settowidth\labelwidth{[999]}
        \leftmargin\labelwidth
        \advance\leftmargin\labelsep
        \@nmbrlisttrue\def\@listctr{pubctr}
        \setcounter{pubctr}{#1}\addtocounter{pubctr}{-1}}}
\def\@@publist{\list
        {[\arabic{pubctr}]\hfill}{\settowidth\labelwidth{[999]}
        \leftmargin\labelwidth
        \advance\leftmargin\labelsep
        \@nmbrlisttrue\def\@listctr{pubctr}}}
 \relax
\makeatother
\catcode`\@=11
\def\section{\@startsection {section}{1}{0pt}{-3.5ex plus -1ex minus
 -.2ex}{2.3ex plus .2ex}{\raggedright\large\bf}}
\catcode`\@=12
\newskip\humongous \humongous=0pt plus 1000pt minus 1000pt

\newif\ifdtup

\def\oldreffmt#1{\rlap{[#1]} \hbox to 2\parindent{}}

\def\figfmt#1{\rlap{Figure {#1}} \hbox to 1in{}}

\def\beq{\begin{equation}}
\def\eeq{\end{equation}}

\def\bea{\begin{eqnarray}}

\def\eea{\end{eqnarray}}
\catcode`\@=11
\def\eqnarray{\stepcounter{equation}\let\@currentlabel=\theequation
\global\@eqnswtrue
\global\@eqcnt\z@\tabskip\@centering\let\\=\@eqncr
\gdef\@@fix{}\def\eqno##1{\gdef\@@fix{##1}}%
$$\halign to \displaywidth\bgroup\@eqnsel\hskip\@centering
  $\displaystyle\tabskip\z@{##}$&\global\@eqcnt\@ne
  \hskip 2\arraycolsep \hfil${##}$\hfil
  &\global\@eqcnt\tw@ \hskip 2\arraycolsep $\displaystyle\tabskip\z@{##}$\hfil
   \tabskip\@centering&\llap{##}\tabskip\z@\cr}

\def\@@eqncr{\let\@tempa\relax
    \ifcase\@eqcnt \def\@tempa{& & &}\or \def\@tempa{& &}
      \else \def\@tempa{&}\fi
     \@tempa \if@eqnsw\@eqnnum\stepcounter{equation}\else\@@fix\gdef\@@fix{}\fi
     \global\@eqnswtrue\global\@eqcnt\z@\cr}

\catcode`\@=12
\font\tenbifull=cmmib10 
\font\tenbimed=cmmib10 scaled 800
\font\tenbismall=cmmib10 scaled 666
\relax

\def\thefootnote{\fnsymbol{footnote}}
\newcommand{\be}{\begin{equation}}
\newcommand{\ee}{\end{equation}}
\newcommand{\bq}{\begin{eqnarray}}
\newcommand{\eq}{\end{eqnarray}}
\newcommand{\g}{\mbox{$(g^{-1}\partial_{+}g)$}}
\newcommand{\h}{\mbox{$(h^{-1}\partial_{-}h)$}}
\begin{document}
\begin{titlepage}
\begin{center}
\today     \hfill    LBL-36926 \\
          \hfill    UCB-PTH-95/07 \\

\vskip .5in

{\large \bf The Conformal Points Of The Generalized Thirring Model II}
\footnote{This work was supported in part by the Director, Office of
Energy Research, Office of High Energy and Nuclear Physics, Division of
High Energy Physics of the U.S. Department of Energy under Contract
DE-AC03-76SF00098 and in part by the National Science Foundation under
grant PHY-90-21139.}

\vskip .5in
Korkut Bardakci and Luis M. Bernardo\footnote{Supported by JNICT (Lisbon).}
\vskip .5in

{\em Theoretical Physics Group\\
    Lawrence Berkeley Laboratory\\
      University of California\\
    Berkeley, California 94720}
\end{center}

\vskip .5in

\begin{abstract}

In the large $N$ limit, conditions for the conformal invariance of the
generalized Thirring model are derived, using two different approaches: the
background field method and the Hamiltonian method based on an operator
algebra, and the agreement between them is established. A free field
representation of the relevant algebra is presented, and the structure of the
stress tensor in terms of free fields (and free currents) is studied in detail.
\end{abstract}
\end{titlepage}
\newpage
\renewcommand{\thepage}{\arabic{page}}
\setcounter{page}{1}
\noindent {\bf 1. Introduction}
\vskip 9pt
In searching for new conformal field theories in two dimensions, a hitherto
relatively less explored candidate is the generalized Thirring model. By
generalized Thirring model, we mean a model of several massless fermions
interacting through the most general Lorentz invariant four fermion couplings,
including parity violating interactions. This is a further generalization
of the parity invariant version  considered in references [1] and [2].
In this paper, which is a sequel to [1], we will continue the investigation
of this more general version of the Thirring model.
The model  is  classically scale invariant , and although scale
invariance is in general broken quantum mechanically, the hope
is that there are isolated points in the coupling constant space where the
 invariance is restored. Since any local conformal field theory
in two dimensions can serve as the basis for  string compactification,
 the construction of new conformal theories of
this type, apart from its own intrinsic interest, can lead to advances in
string theory. Another possible area of application is the statistical
mechanics of two dimensional systems.

A well known and somewhat trivial example of a conformal theory of this type
 is the original Thirring
model [3], which is equivalent to a free field theory. A much less trivial
example is the non-Abelian Thirring model, when the four fermion interaction
is invariant under some Lie group. In a fundamental paper, Dashen and Frishman
[4] showed that this model has conformal invariance at quantized values of the
coupling constant, and that the stress tensor at the conformal points is given
by the affine Sugawara construction [5].
 Unfortunately, much less is known about
the model when the coupling constants are not restricted by any symmetry. There
is some evidence that [6] a model of this type may describe the world sheet of
the string theory resulting from QCD, and if this indeed the case, it is
important to learn more about possible conformal points in the coupling
constant space.
 In the absence of exact solutions of the
Dashen-Frishman type, recent investigations of this model treated the problem
in the large $N$ expansion, $N$ being the number of fermions, and conditions on
the coupling constants in the first non-trivial order in $1/N$ were derived
[1,2,7]. If these results continue to hold in higher orders in $1/N$, the
generalized Thirring model does indeed have conformal points in the coupling
constant space.

In this paper, we will adress several questions related to conformal
invariance of the generalized Thirring model in the large $N$ limit.
One of our aims   is to compare both the methods and the
conclusions of references [1] and [2]. We were
lead to this reexamination because we found that the conditions on the
coupling constants derived in these two references seemed to disagree. Since
the methods used in these two papers are different, it seemed of interest to us
to resolve this conflict. Although both papers start with bosonization, [2]
uses the standard background field approach to examine conformal invariance,
whereas [1] instead uses operator methods and the
Hamiltonian picture in the light cone variables. We feel that it is worthwhile
to supplement the Lagrangian approach with  operator methods based on a
Hamiltonian in order to learn more about the model, and so it is
important to resolve possible conflicts between the two complementary
methods. In section 2, we rederive the
condition for conformal invariance, using   the background field method. Our
calculation is somewhat different from that of [2], and it serves as
a good check on the results of this reference, since we use a
different bosonization scheme which avoids the introduction of the dilaton.
In the end, the conditions we derive turn out to be identical to the
conditions derived in [2] for the case of  parity conserving coupling
constants.
We also notice that these conditions are invariant
 under a set of transformations of
the matrix representing the coupling constants. They consist of an inversion
and multiplications by orthogonal matrices, and they remind us of a similar set
of transformations encountered in toroidal compactification [8].

 The confirmation of the results of reference [2]
 makes it clear that there must be something
wrong with the conditions derived by operator methods in [1]. In section 3, we
reexamine the operator approach and in particular the construction of the
stress tensor. In a conformal theory, the stress tensor should be traceless
and should satisfy the Virasoro algebra. Instead, we find an operator anomaly
in the stress tensor which is equivalent to the well known trace anomaly. This
anomaly, which was missed in [1], was the source of the disagreement
between the operator and background field methods. By imposing the requirement
that this anomaly should vanish, we derive conditions on the coupling constants
in full agreement with the background field method. We also give the operator
construction of the two chiral components of the stress tensor, show that they
satisfy the Virasoro algebra without invoking any further conditions on the
coupling constants, and we compute the two (left,right) central charges.
 This then corrects
and extends to parity non-invariant case the results of [1].

Another aim of this paper is to study further the operator algebra introduced
in [1], which resulted from the quantization of the bosonized Thirring model.
This is of some interest, since this algebra, which can be regarded as a
generalization of the affine Lie algebra, to our best knowledge is new.
 In section 4,
we present a (non-local) representation of this algebra in terms of free
fields, incidentally establishing its consistency beyond any doubt.
One reason for doing this is to see whether the model at hand  can be
mapped into a well-known conformal theory. In particular, we have in mind
the free field theory, and less trivially, the affine Sugawara construction
[5]. The mapping into free fields is non-local and complicated; however,
there is always the hope that the stress tensor may turn out to be something
simple and recognizable. Indeed,the
expression for the stress tensor in terms of the free fields turns out to be
 quadratic, which is an unexpectedly simple result.
 However, in this expression there is an unusual term in which the
second derivative of the free field appears. This term is responsible for the
difference of the central charge from the free field value and it cannot
be eliminated. In a similar fashion, one can reexpress the algebra in terms
of currents that satisfy an affine Lie algebra,
in  the hope that the stress tensor may then admit an affine Sugawara
construction [5]. The stress tensor is indeed quadratic in currents, however,
there is again the term with two derivatives, which is not present in the
affine Sugawara construction. As a result, the affine Sugawara construction,
at least in its simplest form, does not work [9],
 unless additional conditions that go beyond requiring conformal
invariance are imposed. This strongly suggests the emergence of a new conformal
structure. Finally, the last section summarizes our conclusions
and lists problems that await future investigation.
\vskip 9pt
\noindent {\bf 2. Background Field Method and Conformal Invariance}
\vskip 9pt

In this section, using bosonization and the background field method,
 we will investigate  conformal invariance of the generalized
parity non-invariant Thirring model. Our bosonization is based directly on the
Polyakov-Wiegmann [10] method, whereas Tseytlin [2] used
 an approach based on the gauging of the WZW model [11]. In this latter
approach, one has to integrate  over a gauge field, and the non-trivial
integration measure requires the introduction of a dilaton field [12]. In
contrast, we avoid this complication [13]. Our starting point is the parity
violating generalized Thirring model given by
\be
S_{o}=\int d^{2}x(\bar{\Psi}i\gamma^{\mu}\partial_{\mu}\Psi-\tilde{G}_{ab}^{-1}
\; \bar{\Psi}_{R}\lambda_{a}\Psi_{R}\;\bar{\Psi}_{L}\lambda_{b}\Psi_{L})
\ee
where $R$ and $L$ refer to the right and left chiral
 components of $\Psi$, and
the coupling constant $G_{ab}$ is not necessarily symmetric, resulting in
parity violation.
 Upon bosonization [10,1,7], this  gives \footnote{The metric in group
 space is just
$\delta_{ab}$ and so there is no distinction between upper and lower indices.}
\be
S_{o}= W(g)+W(h^{-1})
-\frac{N}{2\pi}\int d^{2}x G_{ab}(ig^{-1}\partial_{+}g)_{a}(ih^{-1}
\partial_{-}h)_{b}
\ee
where $X_{a}$ stands for $Tr(\lambda_{a}X)$ and
\be
G_{ab}=\frac{1}{2}\delta_{ab}-\frac{\pi}{2N}\tilde{G}_{ab}
\ee
and W is the WZW action
\be
W(g)=\frac{N}{8\pi}\left( \int d^{2}x Tr(\partial_{\mu}g^{-1}
\partial^{\mu}g)+\frac{2}{3}\int Tr\left( (g^{-1}dg)^{3} \right)\right).
\ee
Here, we have assumed that $(2G-1)$ is an invertible matrix.
In the absence of sources, the equations of motion are equivalent to
conservation of two currents:
\be
\partial_{+}J_{-}=\partial_{-}J_{+}=0,
\ee
where
\bq
J_{+} & = & i\frac{N}{4\pi}\left(-\partial_{+}h h^{-1}+2 h\lambda_{a}h^{-1}
G_{ba}(g^{-1}\partial_{+}g)_{b}\right), \nonumber \\
J_{-} & = & i\frac{N}{4\pi}\left(-\partial_{-}g g^{-1}+2 g\lambda_{a}g^{-1}
G_{ab}(h^{-1}\partial_{-}h)_{b}\right).
\eq
Two implement the background field method, we add a term to the action
which represents the coupling of two external sources $K_{+,-}$ to two
suitable currents:
\be
\Delta S = \frac{N}{2\pi}\int d^{2}x Tr\left( K_{+}(ih^{-1}\partial_{-}h)
\right) +\frac{N}{2\pi}\int d^{2}x Tr\left( K_{-}(ig^{-1}\partial_{+}g)\right).
\ee
The next step is to define classical fields by solving the equations of
motion in the presence of sources.
 A special solution we are going to use is
\bq
K_{-,a} & = & \left( -\frac{1}{2}(ig^{-1}\partial_{-}g)_{a} + G_{ab}
(ih^{-1}\partial_{-}h)_{b} \right)_{classical} \nonumber \\
K_{+,a} & = & \left( -\frac{1}{2}(ih^{-1}\partial_{+}h)_{a} + G_{ba}
(ig^{-1}\partial_{+}g)_{b} \right)_{classical}
\eq
These sources $K_{+,-}$ can be substituted back in  $S$ to give the
classical action $S^{(0)}$. This defines the classical (background)
fields $g_{clas.}$ and $h_{clas.}$ around which we expand the full quantum
fields $g$ and $h$.
In the appendix, we use background field perturbation theory  to
derive, to one loop order,   the conditions that the coupling constants
$G_{ab}$ must satisfy to have conformal invariance. This is done by
first expanding the action $S$ around $S^{(0)}$, and by calculating the
one loop divergent contribution to the action. This calculation is fairly
standard [14], and for the sake of completeness, it is sketched
in the appendix. The result is
\be
S[\phi] = S^{(0)}[\phi_{clas.}] + S^{(2)}[\phi_{clas.}]
\ee
where $S^{(2)}$ is logarithmically divergent. Here $\phi$ (which stands for
both $\phi$ and $\overline{\phi}$) is the field used
to parametrize $g$ (and $h$ is parametrized by $\overline{\phi}$), and
$\phi_{clas.}$ is defined by $g_{clas.}=
g(\phi_{clas.})$. The divergent piece can be written as (from now on $\phi$
stands for $\phi_{clas.}$)
\be
S^{(2)}[\phi] \cong  \int \frac{d^{2}p}{p^{2}-m^{2}} \int d^{2}x O(x)
\ee
where
\bq
O(x) & = & Y^{(11)}_{ab}E^{a}_{\alpha}E^{b}_{\beta}\partial_{+}\phi^{\alpha}
	\partial_{-}\phi^{\beta} +Y^{(22)}_{ab}\overline{E}^{a}_{\alpha}
	\overline{E}^{b}_{\beta}\partial_{+}\overline{\phi}^{\alpha}
        \partial_{-}\overline{\phi}^{\beta} \nonumber \\
     &   & +Y^{(21)}_{ab} \overline{E}^{a}_{\alpha}E^{b}_{\beta}
	\partial_{+}\overline{\phi}^{\alpha}\partial_{-}\phi^{\beta}
      +Y^{(12)}_{ab}E^{a}_{\alpha}\overline{E}^{b}_{\beta}
	\partial_{+}\phi^{\alpha}\partial_{-}\overline{\phi}^{\beta}
\eq
and
\begin{eqnarray*}
Y^{(11)}_{ab}  =  Tr[-4GH^{-1}G^{T}f_{a}\tilde{H}^{-1}f_{b}] & &
Y^{(22)}_{ab}  =  Tr[-4H^{-1}f_{a}G^{T}\tilde{H}^{-1}Gf_{b}] \\
Y^{(21)}_{ab}  =  Tr[4GH^{-1}f_{a}G^{T}\tilde{H}^{-1}f_{b}] & &
Y^{(12)}_{ab}  =  Tr[4H^{-1}G^{T}f_{a}\tilde{H}^{-1}Gf_{b}]
\end{eqnarray*}
with
$$
H=1-4G^{T}G, \;\;\; \tilde{H}=1-4GG^{T}
$$
and the matrices $f_{a}$ are defined by $(f_{a})_{bc}=f_{abc}$, where $f_{abc}$
are the structure constants of the group. The $E^{a}_{\alpha}$'s are the
 vielbeins defined by
$$
E^{a}_{\alpha}(\phi)\partial_{+}\phi^{\alpha} \equiv Tr(\lambda^{a}ig^{-1}
	\partial_{\alpha}g)\partial_{+}\phi^{\alpha}=Tr(\lambda^{a}ig^{-1}
	\partial_{+}g)
$$
with similar definitions for $\overline{E}^{a}_{\alpha}$'s in terms of $h$'s.
Now compare this  divergent piece with the original Lagrangian, expressed
in terms of classical fields
\bq
S^{(0)} & = & W(g)+W(h^{-1})+\frac{N}{2\pi}\int d^{2}x
G_{ab}E^{a}_{\alpha}\overline{E}^{b}_{\beta}
    \partial_{+}\phi^{\alpha}\partial_{-}\overline{\phi}^{\beta} \nonumber \\
 & & -\frac{N}{4\pi}\int
d^{2}xE^{a}_{\alpha}E_{a\beta}\partial_{+}\phi^{\alpha}
    \partial_{-}\phi^{\beta}-\frac{N}{4\pi}\int
d^{2}x\overline{E}^{a}_{\alpha}\overline{E}_{a\beta}
    \partial_{+}\overline{\phi}^{\alpha}\partial_{-}\overline{\phi}^{\beta}
\eq
 Of the four distinct divergent terms defined by eq.(11), three correspond to
wave function renomalizations and can be eliminated by field redefinitions.
Conformal invariance is then imposed by requiring that the remaining
divergence (the beta function) vanish. The field redefinitions that eliminate
the spurious divergences are given by
\bq
(ig^{-1}\partial_{+}g)_{a} & \longrightarrow & (ig^{-1}\partial_{+}g)_{a} +
\lambda^{(11)}_{ab}(ig^{-1}\partial_{+}g)_{b} + \lambda^{(12)}_{ab}(ih^{-1}
\partial_{+}h)_{b},\nonumber \\
(ih^{-1}\partial_{-}h)_{a} & \longrightarrow & (ih^{-1}\partial_{-}h)_{a} +
\lambda^{(21)}_{ab}(ih^{-1}\partial_{-}h)_{b} + \lambda^{(22)}_{ab}
(ig^{-1}\partial_{-}g)_{b}
\eq
where the $\lambda$'s are first order in $1/N$. This corresponds, in the
Polyakov-Wieg\-mann  bosonization, to making the identification
\bq
A_{+,a} & = & (\delta_{ab} + \lambda^{(11)}_{ab})(ig^{-1}\partial_{+}g)_{b} +
    \lambda^{(12)}_{ab}(ih^{-1}\partial_{+}h)_{b},\nonumber \\
A_{-,a} & = & (\delta_{ab} + \lambda^{(22)}_{ab})(ih^{-1}\partial_{-}h)_{b} +
    \lambda^{(21)}_{ab}(ig^{-1}\partial_{-}g)_{b},
\eq
instead of
$$
A_{+,a} = (ig^{-1}\partial_{+}g)_{a},\;\;A_{-,a} = (ih^{-1}\partial_{-}h)_{a}
$$
The same result can be obtained
by introducing additional sources $L_{+,-}$,
\be
\Delta S = \frac{N}{2\pi}\int d^{2}x Tr\left( L_{-}(ih^{-1}\partial_{+}h)
\right) +\frac{N}{2\pi}\int d^{2}x Tr\left( L_{+}(ig^{-1}\partial_{-}g)\right)
\ee
which are zero to lowest order, and by transforming $K_{+,-}$, $L_{+,-}$
linearly among themselves (source renormalization).
\newline Under these field redefinitions the first order correction
 to $S^{(0)}[\phi]$ is
\bq
\Delta S^{(0)} & = & -\frac{N}{4\pi}\int d^{2}x \left((\lambda^{(21)}_{ab} +
\lambda^{12}_{ba})
\overline{E}^{a}_{\alpha}E^{b}_{\beta}\partial_{+}
    \overline{\phi}^{\alpha}\partial_{-}\phi^{\beta}  \right. \nonumber \\
 &  &  -2(\lambda^{(11)}_{ca}G_{cb}+\lambda^{(22)}_{cb}G_{ac})
E^{a}_{\alpha}\overline{E}^{b}_{\beta}\partial_{+}\phi^{\alpha}\partial_{-}
    \overline{\phi}^{\beta}  \nonumber \\
  &  & +  (\lambda^{(11)}_{ba} - 2G_{ac}\lambda^{(21)}_{cb})E^{a}_{\alpha}
E^{b}_{\beta}\partial_{+}\phi^{\alpha}\partial_{-}\phi^{\beta}\nonumber \\
  & & \left.  +(\lambda^{(22)}_{ab} -2G_{cb}\lambda^{(12)}_{ca})
    \overline{E}^{a}_{\alpha}\overline{E}^{b}_{\beta}\partial_{+}
    \overline{\phi} ^{\alpha}
    \partial_{-}\overline{\phi}^{\beta} \right).
\eq
We now try to eliminate the divergent terms, and this leads to the matrix
 equations
\bq
-Y^{(11)}+\frac{N}{4\pi}((\lambda^{(11)})^{T}-2G\lambda^{(21)}) & = & 0,
    \nonumber \\
-Y^{(12)}-\frac{N}{4\pi}(2(\lambda^{(11)})^{T}G +2G\lambda^{(22)}) & = & 0,
    \nonumber \\
-Y^{(22)}+\frac{N}{4\pi}(\lambda^{(22)} -2(\lambda^{(12)})^{T}G) & = & 0,
    \nonumber \\
-Y^{(21)}+\frac{N}{4\pi}(\lambda^{(21)} +(\lambda^{(12)})^{T}) & = & 0
\eq
At first, one might think that these equations can be solved for the unknown
$\lambda$'s. If this were true, all the infinities would be absorbed by field
redefinitions and conformal invariance would be automatic!
 In fact, the equations are linearly dependent, and for a
solution to exist, the Y's must satisfy the following condition:
$$
Y^{(12)}+2Y^{(11)}G+2GY^{(22)}+4GY^{(21)}G=0.
$$
This  condition is therefore equivalent to the vanishing of the beta function.
Written out explicitly, this leads to the following equation between
the coupling constants:
\bq
Tr[H^{-1}G^{T}f_{a}\tilde{H}^{-1}Gf_{b}]
+4G_{aa'}G_{b'b}Tr[GH^{-1}f_{a'}G^{T}\tilde{H}^{-1}f_{b'}] & & \nonumber \\
-2G_{aa'}Tr[H^{-1}f_{a'}G^{T}\tilde{H}^{-1}Gf_{b}]
-2G_{b'b}Tr[GH^{-1}G^{T}f_{a}\tilde{H}^{-1}f_{b'}] & = & 0,
\eq
Eq.(18) is therefore the condition that determines the conformal points
 in the coupling
constant space. For $G=G^{T}$, it agrees with the result obtained in [2],
 where the $Q$ defined there is related to our $G$ by
\be
Q=2(I-G).
\ee
We end this section by noticing that
this equation is invariant under $(2G) \rightarrow (2G)^{-1}$ and under
$G \rightarrow O^{T}_{1}GO_{2}$, where $O_{1}$ and $O_{2}$ are orthogonal
transformations generated by rotations in group space.
 The first one is the standard duality transformation [8],
already noticed in a classical context in [15]. The second set of
transformations are generated by independent group rotations of left and
right fermions:
\be
\Psi_{R}\rightarrow U_{R}\Psi_{R},\; \Psi_{L}\rightarrow U_{L}\Psi_{L}.
\ee

\vskip 9pt
\noindent {\bf 3. The OPE and Conformal Invariance}
\vskip 9pt

In this section, we shall reexamine the conformal invariance of the theory
from the operator point of view, and show that again the same result as in
the last section (eq.(18)) is obtained, reconciling the background field and
operator methods. Our criterion for conformal invariance is the existence of
a chirally conserved stress tensor: it is well known that this is equivalent
to the absence of the trace anomaly in the stress tensor[16]. Our approach
will be to solve the equations of motion for the quantized fields as a
power series in $1/N$, and
then use this result to construct the stress tensor explicitly.
 We will then see that there is an anomalous term which
violates chiral conservation.
 Conformal invariance is restaured by demanding that
this term vanish, and the resulting condition on the coupling constants
agrees with the result derived in the last section using the background field
method. Before discussing the quantum mechanical complications, we will first
briefly review the classical situation. The two chiral components of the
classical stress tensor, defined by
\bq
T(t,x)=\frac{\pi}{\alpha^{2}}M_{a}(t,x)M_{a}(t,x) \nonumber \\
\tilde{T}(t,x)=\frac{\pi}{\alpha^{2}}N_{a}(t,x)N_{a}(t,x)
\eq
where $\alpha=(4\pi/N)^{1/2}$ , $t\equiv x_{+},x\equiv x_{-}$, and
\bq
M_{a}=(H^\frac{1}{2})_{ab}(ih^{-1}\partial_{x}h)_{b}, & H=1-4G^{T}G,
 \nonumber \\
N_{a}=(\tilde{H}^\frac{1}{2}(2G^{T})^{-1})_{ab}(ih^{-1}\partial_{t}h)_{b},
 & \tilde{H}=1-4GG^{T},
\eq
satisfy the conservation laws
\be
\begin{array}{ccc}
\partial_{t}T(t,x)=0, & \longrightarrow & T_{-}(x)=T(t,x),  \\
\partial_{x}\tilde{T}(t,x)=0, & \longrightarrow & T_{+}(t)=\tilde{T}(t,x),
\end{array}
\ee
and  also satisfy the classical (without central charge) Virasoro
algebra [1]:
\be
T(x)T(y)\cong\frac{1}{(x-y)^{2}}(T(x)+T(y)).
\ee
Now, in the quantum version of the stress tensor we replace
the classical expression by (we will work with the $M_{a}(t,x)$'s, but the same
applies to the $N_{a}(t,x)$'s),
\be
T(t,x)= \frac{\pi}{\alpha^{2}}\lim_{y,u \rightarrow x,t}
(C_{ab}M_{a}(t,x)M_{b}(u,y)-\mbox{sing.terms})
\ee
where $C_{ab}$ is a constant matrix which starts with classical value
$\delta_{ab}$, and has higher order corrections given by
\be
C_{ab}=\delta_{ab} + \sum^{\infty}_{n=2} \alpha^{n} C^{(n)}_{ab}
\ee
due to renormalization. In reference [1], $C_{ab}$ was incorrectly set equal
to $\delta_{ab}$ to all orders in $\alpha$; here,
 we will determine it by requiring that the stress
tensor $T$ satisfy the Virasoro algebra. To do this, and to find the
singular terms to be subtracted, we need the OPE's (operator product expansion)
 between two $M$'s. So we will expand
\be
M_{a}(t,x)=\sum^{\infty}_{n=0} \alpha^{n} M^{(n)}_{a}(t,x)
\ee
and carry calculations up to second order. The strategy for computing OPE's
is the following. We first define $M^{(n)}$'s at a fixed $t$, say $t=0$:
$M^{(n)}_{a}(x)\equiv \mbox{$M^{(n)}_{a}(t=0,x)$}$. The OPE's depend only
on $x$ and they can be deduced from the Poisson brackets at fixed $t$ [1].
The Poisson brackets between the $M$'s and the OPE's that follow from them were
computed in [1]; here we simply take over those results, generalizing them
slightly to take into account of the fact that G is no longer self transpose:
\bq
M_{a}^{(0)}(x) M_{b}^{(0)}(y)& \cong & - \frac{1}{2 \pi (x-y)^{2}} \delta_{ab},
	\nonumber \\
\sum_{n=0}^{1} M_{a}^{(n)}(x) M_{b}^{(1-n)}(y)& \cong & -\frac{1}
    {4 \pi (x-y)} F_{abc} \left(M_{c}^{(0)}(x)+M_{c}^{(0)}(y)\right),
 \nonumber \\
\sum_{n=0}^{2} M_{a}^{(n)}(x) M_{b}^{(2-n)}(y) & \cong & -\frac{1}
{4 \pi (x-y)} F_{abc} \left(M_{c}^{(1)}(x)+M_{c}^{(1)}(y)\right) \nonumber \\
 & & + \frac{1}{2 \pi} E_{ab,a'b'} \log(x-y) M_{a'}^{(0)}(x) M_{b'}^{(0)}(y),
\eq
where the constants $A_{abc}$ and $F_{abc}$ are defined by
\bq
A_{abc} & =& - 2H^{-\frac{1}{2}}_{aa'} H^{-\frac{1}{2}}_{bb'}
        (\tilde{H}^{-\frac{1}{2}} G)_{cc'} f_{a'b'c'}
 	+4(H^{-\frac{1}{2}}G^{T})_{aa'}
        (H^{-\frac{1}{2}} G^{T})_{bb'} \tilde{H}^{-\frac{1}{2}}_{cc'}
f_{a'b'c'}
 	,\nonumber \\
F_{abc}&  =& H^{-\frac{1}{2}}_{aa'} H^{-\frac{1}{2}}_{bb'}
	H^{-\frac{1}{2}}_{cc'} f_{a'b'c'}
 	- 8 ( H^{-\frac{1}{2}} G^{T})_{aa'}
	(H^{-\frac{1}{2}} G^{T})_{bb'} (H^{-\frac{1}{2}} G^{T})_{cc'}
	f_{a'b'c'}, \nonumber
\eq
and
\be
E_{ab,cd}= A_{cae} A_{bde}.
\ee
The $N^{(n)}_{a}$'s obey similar OPE's, obtained from the above ones under
$G \rightarrow G^{T}$, with the new constants $\tilde{A}_{abc}$ and
$\tilde{F}_{abc}$,
\bq
\tilde{A}_{abc} & =& - 2 \tilde{H}^{-\frac{1}{2}}_{aa'}
    \tilde{H}^{-\frac{1}{2}}_{bb'}
        (H^{-\frac{1}{2}} G^{T})_{cc'} f_{a'b'c'}+
	4 (\tilde{H}^{-\frac{1}{2}} G)_{aa'}
        (\tilde{H}^{-\frac{1}{2}} G)_{bb'} H^{-\frac{1}{2}}_{cc'} f_{a'b'c'},
	 \nonumber \\
\tilde{F}_{abc}&  =& \tilde{H}^{-\frac{1}{2}}_{aa'}
 	\tilde{H}^{-\frac{1}{2}}_{bb'}
	\tilde{H}^{-\frac{1}{2}}_{cc'} f_{a'b'c'}
 	- 8 ( \tilde{H}^{-\frac{1}{2}} G)_{aa'}
	(\tilde{H}^{-\frac{1}{2}} G)_{bb'}
	(\tilde{H}^{-\frac{1}{2}} G)_{cc'}
	f_{a'b'c'}, \nonumber
\eq
and
\be
\tilde{E}_{ab,cd}= \tilde{A}_{cae} \tilde{A}_{bde}.
\ee
To extend these OPE's to $t\neq 0$, we solve the equations of motion
up to second order, and express the $M$'s and $N$'s at arbitrary $t$ in terms
of the same variables at $t=0$. Since the OPE's at $t=0$ are already known,
they are then easily extended to $t\neq 0$. From
\be
g^{-1}(\partial_{+}J_{-})g=0
\ee
we have
\be
-\partial_{-}\g +2[\g,\lambda_{a}]G_{ab}\h_{b} + 2\lambda_{a}G_{ab}
\partial_{+}\h_{b}=0.
\ee
Now solve for \g\ in terms of $J_{+}$,
\be
\g_{a}=(2G^{T})^{-1}_{ab}(h^{-1}(\partial_{+}-\frac{4i\pi}{N}J_{+})h)_{b}.
\ee
The model is invariant under the gauge transformations  $h\rightarrow u_{+}
(x_{+})h$;
using this gauge invariance , we can set
$J_{+}=0$ ($\partial_{-}J_{+}=0$, so $J_{+}$ depends only on $x_{+}=t$).
It is amusing to notice that the equations of motion can then be written
as flatness conditions for two vector fields $V$ and $W$:
\bq
\partial_{+}V_{-}\;-\partial_{-}V_{+}\;-i[V_{+}\;,V_{-}\;] & = & 0, \nonumber
\\
\partial_{+}W_{-}-\partial_{-}W_{+}-i[W_{+},W_{-}] & = & 0,
\eq
where,
\be
V_{\pm ,a}  =  (ih^{-1}\partial_{\pm}h)_{a}, \nonumber
\ee
\be
W_{+,a}  = (2G^{T})^{-1}_{ab}V_{+,b},\;\;\; W_{-,a}  =  (2G)_{ab}V_{-,b}.
\ee
These can be cast in a more useful form in terms of
\be
M_{a}  = (H^{\frac{1}{2}})_{ab}V_{-,b}, \;\;\;
N_{a}  = (\tilde{H}^{\frac{1}{2}})_{ab}W_{+,b}
\ee
defined before. The equations of motion are then
\bq
\partial_{t}M_{a} & = & -\alpha A_{abc}M_{b}N_{c},  \\
\partial_{x}N_{a} & = & -\alpha \tilde{A}_{abc}N_{b}M_{c}.
\eq
The conservation laws of the (classical) stress tensors follow at once from
these equations due to the antisymmetry of $A_{abc}$ and $\tilde{A}_{abc}$ in
the first two indices.
\newline
The next step  is to  solve the equations of motion iteratively, using the
 expansion in $\alpha$ (eq.(27)), and a similar expansion for $N$.
\newline The zeroth and first order  solutions are
\bq
M^{(0)}_{a}(t,x) & = & M^{(0)}_{a}(x), \nonumber \\
N^{(0)}_{a}(t,x) & = & N^{(0)}_{a}(t). \\
M^{(1)}_{a}(t,x) & = & M^{(1)}_{a}(x) -A_{abc}M^{(0)}_{b}(x)\int^{t}dt'
	N^{(0)}_{c}(t'), \nonumber  \\
N^{(1)}_{a}(t,x) & = & N^{(1)}_{a}(t) -\tilde{A}_{abc}N^{(0)}_{b}(t)\int^{x}dx'
	M^{(0)}_{c}(x'),
\eq
and to second order
\bq
M^{(2)}_{a}(t,x) & = & M^{(2)}_{a}(x) -A_{abc}\int^{t}dt'M^{(0)}_{b}(x)
	N^{(1)}_{c}(t')  \nonumber \\
 & &	+A_{abc}\tilde{A}_{cde}\int^{x}dx'\int^{t}dt'
M^{(0)}_{b}(x)M^{(0)}_{e}(x')
         N^{(0)}_{d}(t')  \nonumber \\
 & &	-A_{abc}\int^{t}dt'M^{(1)}_{b}(x)N^{(0)}_{c}(t')  \\
 & & 	+A_{abc}A_{bde}\int^{t}dt'\int^{t'}dt''M^{(0)}_{d}(x)
	N^{(0)}_{c}(t')N^{(0)}_{e}(t''). \nonumber
\eq
We will not need $N^{(2)}_{a}(t,x)$. Therefore, $M_{a}(t,x)$ can be expressed
in
 terms of
$M^{(n)}_{a}(x)$'s and $N^{(n)}_{a}(t)$'s, functions of only one variable.
If we substitute the above in the definition of $T$ to second order, it is
 easy to see that classically
all of the $t$ dependent terms cancel, as they must, because
we know from the equations of motion that this is true to all orders.
However this does not happen in the quantum case,
 where $M^{(n)}_{a}(x), n=0,1,2$, become operators that satisfy
the OPE's given earlier (eq.(28)). First of all,
as it stands, the above expression for $M_{a}(t,x)$
is not well defined, because we haven't defined yet the product of two or
more $M$'s at the same point. These products
  should be understood as  nonsingular ``normal ordered''
products. For instance, $M^{(0)}_{a}(x)M^{(0)}_{b}(y)$ should be
understood as
\be
:\!\!M^{(0)}_{a}(x)M^{(0)}_{b}(y)\!\!:\,\equiv M^{(0)}_{a}(x)M^{(0)}_{b}(y)
 + \frac{\delta_{ab}}{2\pi(x-y)^{2}}
\ee
and the same applies for the $N^{(n)}_{a}$'s. The product of a $M^{(n)}_{a}$
and a $N^{(n)}_{a}$ gives no problem since they are functions of different
variables and commute with each other.
 This guarantees that \( \lim_{y\rightarrow x}:\!\!M^{(0)}_{a}(x)
M^{(0)}_{b}(y)\!\!: \), and consequently  the above expression for
$M_{a}^{(2)}(t,x)$ is well defined.

 Next, we examine eq.(25), to see what subtractions are needed to make the
stress tensor well-defined, and whether it is $t$ independent,
 as the conservation
law (eq.(23)) demands. It turns out that to the order
 we are considering (second
order in $\alpha$), $T$ can be made finite by making suitable subtractions, and
that all of the terms in $T$, with the possible exception of one term,
 are $t$ independent. The critical term in question,
up to a multiplicative factor of $\pi/\alpha^{2}$, is
$$
T_{critical}  =  A_{abc}\tilde{A}_{cde}\int^{x}dx'
	\int^{t}dt':\!\!M^{(0)}_{b}(x)M^{(0)}_{e}(x')\!\!:
	     \!M^{(0)}_{a}(y)N^{(0)}_{d}(t') +(x \leftrightarrow y)
$$
This term is finite as $y \rightarrow x$ and needs no subtraction. However,
it is $t$ dependent, and therefore, if it does not vanish, it violates the
conservation law (eq.(23)). It does not automatically vanish because,
while $A_{abc}$ is antisymmetric in $a$ and $b$, $:\!\!M^{(0)}_{b}(x)
M^{(0)}_{e}(x')\!\!:\!M^{(0)}_{a}(y) + (x \leftrightarrow y)$ is not symmetric
due to the normal ordering of the two $M$'s. However, the completely normal
ordered product
$$
:\!\!M^{(0)}_{b}(x)M^{(0)}_{e}(x')M^{(0)}_{a}(y)\!\!: + (x \leftrightarrow y)
$$
is symmetric in $a$ and $b$ and vanishes when multiplied by $A_{abc}$. We now
make use of the identity
\bq
\lefteqn{:\!\!M^{(0)}_{b}(x)M^{(0)}_{e}(x')\!\!:\!M^{(0)}_{a}(y)=} \nonumber \\
 & & - \frac{\delta_{ab}}{2\pi(x-y)^{2}}M^{(0)}_{e}(x')
 	- \frac{\delta_{ae}}{2\pi(x'-y)^{2}}M^{(0)}_{b}(x)
	+ :\!\!M^{(0)}_{b}(x)M^{(0)}_{e}(x')M^{(0)}_{a}(y)\!\!: \nonumber
\eq
to find
\bq
\lim_{y\rightarrow x} T_{critical}& = & - \lim_{y\rightarrow x}
	 A_{abc}\tilde{A}_{cde}\int^{t}dt'N^{(0)}_{d}(t') \nonumber \\
      & & \left(\int^{x}dx'\frac{\delta_{ae}}{2\pi(x'-y)^{2}}M^{(0)}_{b}(x)
   	 +\int^{y}dx'\frac{\delta_{ae}}{2\pi(x'-x)^{2}}M^{(0)}_{b}(y)
	\right) \nonumber \\
	& = & -  \lim_{y\rightarrow x} A_{abc}\tilde{A}_{cde} \left(-\frac{1}{2\pi}
	\frac{M^{(0)}_{b}(x)-M^{(0)}_{b}(y)}{x-y}\right)\int^{t}dt'
	N^{(0)}_{d}(t')\nonumber \\
	& = & +\frac{1}{2\pi}A_{bac}\tilde{A}_{dca}M'^{(0)}_{b}(x)\int^{t}dt'
	N^{(0)}_{d}(t') \nonumber
\eq
To eliminate this anomaly and restaure conformal invariance, we have to set its
coefficient equal to zero,
\be
A_{acd}\tilde{A}_{bdc}=0,
\ee
 recovering the same condition as before (eq.(18)). We note that conformal
invariance imposes no restrictions on $C_{ab}^{(2)}$. These constants can be
  determined by requiring that the stress tensor satisfy
the Virasoro  algebra to second order. We therefore need the OPE of the
product of two stress tensors; this was given by eq.(5.5a) of reference [1].
This result has to be modified slightly to take into account that
$C_{ab}^{(2)}$ no longer vanishes. With this modification,
 the OPE of two $T$'s is
\bq
\lefteqn{T(x)T(y) \cong  \frac{c}{2(x-y)^{4}}- \frac{\pi}{(x-y)^{2}}
    \left(M^{(0)}_{a}(x)M^{(0)}_{a}(x)
        +M^{(0)}_{a}(y)M^{(0)}_{a}(y) \right. } \nonumber \\
   & & \left.  + \alpha^{2}M^{(1)}_{a}(x)M^{(1)}_{a}(x)
	+\alpha^{2}M^{(1)}_{a}(y)M^{(1)}_{a}(y)\right)
	-\frac{\alpha^{2}}{4(x-y)^{2}}  \nonumber \\
    & &(F_{aa'b}F_{aa'c}+2E_{aa,bc}+4\pi C^{(2)}_{bc})
        \left(M^{(0)}_{b}(x)M^{(0)}_{c}(x)+M^{(0)}_{b}(y)M^{(0)}_{c}(y) \right)
\nonumber
\eq
where $c$ is the central charge.
\newline Since the Virosoro algebra reads
\be
T(x)T(y)\cong - \frac{1}{(x-y)^{2}}(T(x)+T(y))+\frac{c}{2(x-y)^{4}},
\ee
we must have
\be
F_{aa'b}F_{aa'c}+2E_{aa,bc} +4\pi C^{(2)}_{bc}=0,
\ee
which determines $C^{(2)}_{ab}$, and the central charge $c$ is given by
\be
c=D - \frac{\alpha^{2}}{4\pi}(3E_{aa,bb}+F_{abc}F_{abc}),
\ee
where $D$ is the dimension of the flavor algebra.
This is the central charge of the algebra generated by $T$. The central charge
of the algebra generated by the other chiral component, $\tilde{T}$, (see
eq.(21)) can be gotten from eq.(47), by replacing $E$ by $\tilde{E}$
and $F$ by $\tilde{F}$.

As a check on our formalism, we notice that, at $G_{ab}=0$ in eq.(2),
the action is a sum of two decoupled WZW models and therefore it is
obviously conformal. $G=0$ indeed satisfies the condition for
conformal invariance given by eq.(18) and so the equation passes this
test. There is a further check on the central charge. The stress tensor
of the WZW model is given by the Sugawara construction in terms of the
currents, with the standard formula for the central charge:
\be
c= \frac{2 k D}{2 k + c_{\psi}^{g}},
\ee
where $k$ is the level number of the affine algebra, related to our
$N$ by $2 k=N$ and
$$
c^{g}_{\psi} \delta_{ab}= \sum_{c,d=1}^{D} f_{acd} f_{bcd}.
$$
This formula is exact. We have to compare it with eq.(47) in the
limit of large $N$ (or $k$), with $G$ set equal to zero. In this limit
$E_{ab,cd}=0$, $F_{abc}= f_{abc}$ and so from eq.(47)
$$
c= D \left( 1 - \frac{c_{\psi}^{g}}{N} \right)
$$
which agrees with the standard formula eq.(48) to first order in
$1/N$. This particular solution ($G=0$) has some relation to the
Dashen-Frishman conformal point [4]. It is natural to suspect
such a relation, since both $G=0$ and the Dashen-Frishman
solution are $SU(n)$ symmetric. We do not know how to make
a detailed comparison, except to note that the stress tensor
 of the Dashen-Frishman solution is given by the Sugawara
construction and the central charge is therefore given by
eq.(48). But, as we pointed out above, the $G=0$ solution,
being the sum of two WZW models, has also a Sugawara stress
tensor and the standard formula for the central charge.
Therefore, at the level of stress tensors, there is
agreement.

\vskip 9pt
\noindent {\bf 4. Free Field Realization}
\vskip 9pt

In this section, we will express
 the fields $M^{(n)}_{a}(x)$ in terms of free
fields $\phi_{a}(x)$'s so that the PB in the classical case [1], or the OPE in
the quantum case (eq.(28)), between two $M_{a}(x)$'s is still satisfied.
(These $\phi$'s are not to be confused with the $\phi$'s introduced in section
2). As in the rest of the paper, the calculations will be carried only to
second order in $\alpha$.
 Our motivation for doing this is twofold: first of all, one may ask
whether the relatively complicated appearance of the OPE's given by eq.(28)
is due to our choice of fields; with a different choice of fields, a simpler
algebra might emerge. Indeed, we show that one can express everything in
terms of free fields; however, the simplification achieved in this way is
somewhat illusory, since the expressions connecting $M$'s to free fields
are non-local and complicated. Next, we reexpress the stress tensor in terms
of free fields, hoping for a simple result. Indeed, the stress tensor turns
out to be local and quadratic in free fields; on the other hand, an unusual
term involving the second derivative of the fields makes its appearance.
(The last term in eq.(54)) This
term is responsible for the deviation of the central charge from the free
field value and it cannot be eliminated. Although we will not present the
details here, the $M$'s can also be expressed in terms of currents that
satisfy an affine Lie algebra; in fact, with minor modifications, eqs.(50)
and (53) still hold if the $\phi'_{a}(x)$'s are replaced by currents.
 Again, the stress tensor is quadratic in the currents, as in eq.(54),
which looks promising for an affine Sugawara construction. But again there
appears the analogue of the last term in eq.(54), which, expressed in
terms of the currents $J_{a}(x)$, looks like
$$
J^{\prime}_{a}(x)\int^{x}dy J_{a}(y)
$$
and clearly does not belong in the affine Sugawara construction.

 We start with the classical $M$ fields and try to express them
 in terms of $\phi_{a}(x)$'s that satisfy the free field PB relations,
\be
\{\phi_{a}(x),\phi_{b}(y)\} = -\log(x-y)\delta_{ab}
\ee
The solution  to zeroth order ($M^{(0)}_{a}(x)$)  is obvious, and the next two
orders are easily constructed by guesswork. The result is,
\bq
M^{(0)}_{a}(x) & = & \phi'_{a}(x),  \nonumber \\
M^{(1)}_{a}(x) & = & \frac{1}{3}F_{abc}\phi'_{b}(x)\phi_{c}(x), \nonumber \\
M^{(2)}_{a}(x) & =&  -\frac{1}{36}(F_{ace}F_{bde}+F_{ade}F_{bce})\phi'_{b}(x)
	\phi_{c}(x)\phi_{d}(x)  \nonumber \\
	& &	+\left( \frac{1}{36}F_{abe}F_{cde} +\frac{1}
	{4}E_{ac,bd} \right)\phi'_{b}(x)\int^{x}dy\phi_{c}(y)\phi'_{d}(y).
\eq
This can easily be extended to operators. Define now quantum free
 fields by OPE's
\be
\phi_{a}(x)\phi_{b}(y) \cong - \frac{1}{2\pi}\log(x-y)\delta_{ab}
\ee
To avoid singular expressions we work with normal ordered fields, for
example,
$$
\phi_{a}(x)\phi_{b}(y)\phi_{c}(z)  =
:\!\!\phi_{a}(x)\phi_{b}(y)\phi_{c}(z)\!\!:
	-\frac{1}{2\pi}\phi_{c}(z)\log(x-y)\delta_{ab}
$$
\be
        -\frac{1}{2\pi}\phi_{b}(y)\log(x-z)\delta_{ac}
	-\frac{1}{2\pi}\phi_{a}(x)\log(y-z)\delta_{bc}
\ee
In order to satisfy the OPE algebra given before (eq.(28)),
 we simply take over the classical expression, replacing products
 of fields by normal ordered products. It turns out that this almost
 works; however, additional terms are necessary to make it work.
The final result is
\bq
M^{(0)}_{a}(x) & = & \phi'_{a}(x),  \nonumber \\
M^{(1)}_{a}(x) & = & - \frac{1}{3}F_{abc}:\!\!\phi'_{b}(x)\phi_{c}(x)\!\!:,
 \nonumber \\
M^{(2)}_{a}(x) & = & -\left( \frac{1}{18\pi}F_{acd}F_{bcd} +\frac{1}{4\pi}
		E_{ab,cc}\right) \phi'_{b}(x) \nonumber \\
	& & +\left( \frac{1}{36\pi}F_{acd}F_{bcd}+\frac{1}{4\pi}E_{ab,cc}
\right)
	\int^{x}\frac{dy}{y-x}(\phi'_{b}(y)-\phi'_{b}(x)) \nonumber \\
	& &  -\frac{1}{36}(F_{ace}F_{bde}+F_{ade}F_{bce}):\!\!\phi'_{b}(x)
	\phi_{c}(x)\phi_{d}(x)\!\!:  \nonumber \\
	& &	+\left( \frac{1}{36}F_{abe}F_{cde} +\frac{1}
	{4}E_{ac,bd} \right):\!\!\phi'_{b}(x)\int^{x}dy\phi_{c}(y)
\phi'_{d}(y)\!\!:.
\eq
Using these expressions we can construct the stress tensor, which to second
order, is quadratic in the free fields and is given by
\bq
\lefteqn{\frac{\alpha^{2}}{\pi}T(x) = \; :\!\! \phi'_{a}(x)\phi'_{a}(x) \!\!:}
\nonumber \\
    	& & - \frac{ \alpha^{2}}{24 \pi} \left( F_{acd}F_{bcd} +3
	E_{bc,ac} \right) \left( :\!\! \phi'_{a}(x)\phi'_{b}(x) \!\!:
	+ :\!\! \phi_{a}(x)\phi''_{b}(x) \!\!: \right)
\eq
It can also be directly checked that, at least to second order, this
 construction in terms of free fields yields the Virasoro algebra with
the correct central charge.

\vskip 9pt
\noindent {\bf 5. Conclusions}
\vskip 9pt

The main result of this paper is eq.(18), the condition on the coupling
constants derived  by imposing conformal invariance on the generalized
Thirring model. This result, valid in the large $N$ limit, is obtained by
using two different approaches, the background field method and the
operator method. It corrects and extends the results obtained in [1],
bringing them in agreement with those of [2]. Among the problems that are
still left open is the contribution of the higher order corrections in
$1/N$ to both the condition for conformal invariance (eq.(18)), and to the
operator algebra (eq.(28)).

We have also tried tried to shed some light on the structure of the
operator algebra mentioned above by expressing it in terms of free fields
and free currents. We have found some simplification in the expression for
the stress tensor, but still the result could not be reproduced by any
 well-known  construction. It appears very likely that we have a completely
new conformal model
\vskip 9pt
\noindent {\bf Acknowledgements}
\vskip 9pt

This work was supported in part by the Director, Office of
Energy Research, Office of High Energy and Nuclear Physics, Division of
High Energy Physics of the U.S. Department of Energy under Contract
DE-AC03-76SF00098 and in part by the National Science Foundation under
grant PHY-90-21139.

\vskip 9pt
\noindent{\bf Appendix}
\vskip 9pt

In this appendix, we fill the gaps in the evaluation of
 $S^{(2)}[\phi_{clas.}]$ done in section 2. As explained there, we want
 to expand the action $S[\phi]$ around $S^{(0)}[\phi_{clas.}]$, the
 classical action. To do this, parametrize the fields $g$ and $h$ by:
\be
g=g(\phi),\;\;\; h=h(\overline{\phi}),
\ee
where $\phi(x)$  stands for $\phi^{\alpha}(x)$. The $\phi^{\alpha}$'s are the
coordinates in the group manifold where $g$  takes values,
 and $x\equiv x^{\mu}$
are coordinates in Minkowski 2-space. The classical fields
 $\phi^{\alpha}_{clas.}$
are defined by $g_{clas.}=g(\phi_{clas.})$, and similarly for
$\overline{\phi}^{\alpha}$. From now on, unless otherwise stated, $\phi$
stands either for $\phi$ and $\overline{\phi}$.

 Using the vielbeins $E^{a}_{\alpha}(\phi)$ and $\overline{E}^{a}_
{\alpha}(\overline{\phi})$ defined in section 2, the source terms can
be written as
\bq
&   Tr\left( K^{-}(ig^{-1}\partial_{+}g)\right)=Tr( K^{-}\lambda_{a})
    E^{a}_{\alpha}\partial_{+}\phi^{\alpha}
    \equiv K^{-}_{a}E^{a}_{\alpha}\partial_{+}\phi^{\alpha},
    & \nonumber  \\
&   Tr\left( K^{+}(ih^{-1}\partial_{-}h)\right)=Tr( K^{+}\lambda_{a})
    \overline{E}^{a}_{\alpha}\partial_{-}\overline{\phi}^{\alpha}
    \equiv K^{+}_{a}\overline{E}^{a}_{\alpha}\partial_{-}\overline{\phi}
^{\alpha}, &
\eq
and the action becomes
\bq
S & = & W(g)+W(h^{-1})-\frac{N}{2\pi}\int d^{2}x G_{ab}E^{a}_{\alpha}
    \overline{E}^{b}_{\beta}\partial_{+}\phi^{\alpha}\partial_{-}
    \overline{\phi}^{\beta}  \nonumber \\
 & & +\frac{N}{2\pi}\int d^{2}xK^{+}_{a}\overline{E}^{a}_{\alpha}\partial_{-}
    \overline{\phi}^{\alpha}+\frac{N}{2\pi}\int d^{2}xK^{-}_{a}E^{a}_{\alpha}
    \partial_{+}\phi^{\alpha}.
\eq
Now we expand this action $S[\phi]$ around the classical action
$S^{(0)}=S[\phi_{clas.}]$
treating $K_{+,-}$ as classical sources, which can then be written in terms of
$\phi_{clas.}$.
To expand the action, let
$$
\phi(x)\longrightarrow\phi(x,s)
$$
so that $\phi(x,s=0) = \phi_{clas.}(x)$ and $\phi(x,s=1) = \phi(x)$ and define
\be
\xi^{\alpha}=\frac{d}{ds}\phi^{\alpha}(x,s)\!\mid_{s=0}
\ee
The $\xi^{\alpha}(x)$'s span the tangent space at $\phi_{clas.}(x)$ and satisfy
the geodesic
equation
\be
\frac{D}{Ds} \xi^{\alpha} = \frac{d}{ds}\xi^{\alpha}
 + \Gamma^{\alpha}_{\beta\gamma}\xi^{\beta}\xi^{\gamma} = 0,
\ee
where
\be
\Gamma^{\alpha}_{\beta\gamma} = \frac{1}{2}E^{\alpha}_{a}
 \left( \frac{\partial}{\partial \phi^{\beta}}E^{a}_{\gamma}
 + \frac{\partial}{\partial \phi^{\gamma}}E^{a}_{\beta} \right)
\ee
is the Christoffel symbol. In general the $\xi^{\alpha}(x)$'s don't form
an orthonormal basis but we can define new vectors
\be
\zeta^{a}=E^{a}_{\alpha}\xi^{\alpha}
\ee
that span the tangent space at $\phi_{clas.}(x)$ and form an orthonormal basis.
The inverse relation is given by
\be
\xi^{\alpha}=E_{a}^{\alpha}\zeta^{a}
\ee
where $E_{a}^{\alpha}(\phi)$ is the inverse vielbein defined by
\be
E^{a}_{\alpha}E^{\alpha}_{b}=\delta_{ab}.
\ee
Note that in the $\{\zeta^{a}\}$ basis the metric is $\delta_{ab}$ and so there
is no diference
between upper and lower indices, while in the $\{\xi^{\alpha}\}$ basis the
metric is $g_{\alpha \beta}=
E^{a}_{\alpha}E_{a \beta}$ and so an upper index is different from a lower
index.
The action can then be expanded as
\be
S[\phi(x,s)]\!\mid_{s=1}=\sum_{n=0}^{\infty}\frac{1}{n!}\left( \frac{d}{ds}
\right) ^{n}S[\phi]\!\mid_{s=0}\equiv \sum_{n=0}^{\infty}
S^{(n)}[\phi_{clas.},\zeta],
\ee
and keeping terms to second order in $\zeta$'s we have (from now
on $\phi(x)$ stands for $\phi_{clas.}(x)$):
\bq
S^{(0)}[\phi,\zeta] & =& S[\phi] , \nonumber \\
S^{(1)}[\phi,\zeta] & =& 0  \;\;\;\mbox{         if equations of motion hold}
\nonumber \\
S^{(2)}[\phi,\zeta] & =& \frac{N}{8\pi}\int d^{2}x \left(
	 \zeta^{a}(-\delta_{ab}\Box+ A^{\mu}_{ab}\partial_{\mu} +
 	D_{ab})\zeta^{b} \right.\nonumber \\
	& & +\overline{\zeta}^{a}(-\delta_{ab}\Box+
	\overline{A}^{\mu}_{ab}\partial_{\mu} +
	 \overline{D}_{ab})\overline{\zeta}^{b}  \nonumber \\
	 & &  +\zeta^{a}(2G_{ab}\Box+ B^{\mu}_{ab}\partial_{\mu} +
	 C_{ab})\overline{\zeta}^{b} \nonumber \\
	& & \left.+\overline{\zeta}^{a}(2G_{ba}\Box+
	\overline{B}^{\mu}_{ab}\partial_{\mu} +
	C_{ba})\zeta^{b} \right),
\eq
where
\bq
A^{\mu} & = & 2f_{n}E^{n}_{\alpha}\partial^{\mu}\phi^{\alpha}, \nonumber \\
\overline{A}^{\mu} & = & 2f_{n}\overline{E}^{n}_{\alpha}\partial^{\mu}
\overline{\phi}^{\alpha}, \nonumber \\
B^{\mu} & = & -4f_{n}GE^{n}_{\alpha}(\eta^{\mu\nu}-\varepsilon^{\mu\nu})
\partial_{\nu}\phi^{\alpha}, \nonumber \\
\overline{B}^{\mu} & = & -4f_{n}G^{T}\overline{E}^{n}_{\alpha}
    (\eta^{\mu\nu}+\varepsilon^{\mu\nu})\partial_{\nu}
    \overline{\phi}^{\alpha}, \nonumber \\
C & = & 2f_{n}Gf_{m}E^{n}_{\alpha}\overline{E}^{m}_{\beta}
    (\eta^{\mu\nu}+\varepsilon^{\mu\nu})
    \partial_{\mu}\phi^{\alpha}\partial_{\nu}\overline{\phi}^{\beta},
 \nonumber \\
D & = & -f_{m}f_{n}E^{m}_{\alpha}E^{n}_{\beta}\partial_{\mu}\phi^{\alpha}
    \partial^{\mu}\phi^{\beta}, \nonumber \\
    \overline{D} & = & -f_{m}f_{n}\overline{E}^{m}_{\alpha}
    \overline{E}^{n}_{\beta}\partial_{\mu}\overline{\phi}^{\alpha}
    \partial^{\mu}\overline{\phi}^{\beta},
\eq
and the matrix $f_{n}$ is defined by $(f_{n})_{ab}=f_{nab}$.
To compute the divergent counter term, we write $S^{(2)}$ in the form
\be
S^{(2)}=\frac{N}{8\pi}\int d^{2}x Z^{T}(R\Box + P^{\mu}\partial_{\mu} + Q)Z
\ee
where
\be
\begin{array}{ccc}
    Z\equiv \left[ \begin{array}{c}
        \zeta \\
        \overline{\zeta}
            \end{array} \right]
&
    \mbox{and}
&
    \begin{array}{cc}
    \zeta =\left[ \begin{array}{c}
		\zeta^{1} \\ \vdots\\ \zeta^{n}
		\end{array} \right]
    &
    \overline{\zeta} =\left[ \begin{array}{c}
		\overline{\zeta}^{1} \\ \vdots\\ \overline{\zeta}^{n}
		\end{array} \right]
    \end{array}
\end{array}
\ee
and the matrices $R$, $P^{\mu}$ and $Q$ are
\be
\begin{array}{ccc}
R=\left[ \begin{array}{cc}
		-I & 2G \\
		2G^{T} & -I
		\end{array} \right]
&
P^{\mu}=\left[ \begin{array}{cc}
		A^{\mu} & B^{\mu} \\
		\overline{B}^{\mu} & \overline{A}^{\mu}
		\end{array} \right]
&
Q=\left[ \begin{array}{cc}
		D & C \\
		C^{T} & \overline{D}
		\end{array} \right]
\end{array}
\ee
After integrating over $Z$, we get
\bq
S^{(2)} & \cong & -\frac{1}{2}Tr \log (R\Box +P^{\mu}\partial_{\mu} +Q)
\nonumber \\
    & \cong & -\frac{1}{2}Tr \left(R^{-1}\frac{1}{\Box}Q-\frac{1}{2}R^{-1}
        \frac{1}{\Box}P^{\mu}\partial_{\mu}R^{-1}\frac{1}{\Box}P^{\nu}
        \partial_{\nu} \right)  \nonumber\\
	 & \cong & \int \frac{d^{2}p}{p^{2}-m^{2}} \int d^{2}x O(x),
\eq
where $O(x)$ was defined in section 2.
\newpage
{\bf References}
\begin{enumerate}
\item  K.Bardakci, Nucl.Phys. {\bf B431} (1994) 191.
\item  A.A.Tseytlin, Nucl.Phys. {\bf B418} (1994) 173.
\item  W.Thirring, Ann. Phys. (N.Y.) {\bf 3} (1958) 91.
\item  R.Dashen and Y.Frishman, Phys.Rev. {\bf D11} (1975) 2781.
\item  For a review of the affine Sugawara construction, see ''Irrational
Conformal Field Theory'', UCB-PTH-95/02, hep-th/9501144.
\item  K.Bardakci, Nucl.Phys. {\bf B401}(1993) 168.
\item  C.Hull and O.A.Soloviev, QMW-PH 95-9, hep-th/9503021.
\item  For a review, see A.Giveon, M.Porrati and E.Rabinovici, Phys. Rep.
    {\bf 244} (1994) 77.
\item  See however [5] for a different approach.
\item  A.M.Polyakov and P.B.Wiegmann, Phys. Lett. {\bf B131} (1983) 121,
{\bf B141} (1984) 223.
\item  D.Karabali, Q.H.Park and H.J.Schnitzer, Nucl. Phys. {\bf B 323}
(1989) 572, Phys. Lett. {\bf B205} (1988) 267.
\item  A.A.Tseytlin, Nucl. Phys. {\bf B411} (1994) 509.
\item  For a comparision of the two approaches, see L.S.Brown and
R.I.Nepo\-mechie,
    Phys. Rev., {\bf D35} (1987) 3239.
\item  C.G.Callen, D.Friedan, E.Martinec and M.J.Perry, Nucl. Phys. {\bf B262}
(1985) 593.
\item  O.A.Soloviev, Mod. Phys. Lett. {\bf A8} (1993) 301.
\item  A.Belavin, A.M.Polyakov and A.B.Zamolodchikov, Nucl. Phys. {\bf B 241}
(1984) 333.
\end{enumerate}

\end{document}